\documentclass[preprint]{revtex4}

\usepackage{graphicx}
\newcommand{\sgn}{\,{\rm sgn}\,}

\renewcommand{\vec}[1]{\mbox{\boldmath $#1$}}

\begin{document}

\title{Generating functional analysis of CDMA detection dynamics}
\author{Kazushi Mimura}
\affiliation{
Faculty of Information Sciences, Hiroshima City University, Hiroshima 731-3194, Japan}
\email{mimura@cs.hiroshima-cu.ac.jp}
\author{Masato Okada}
\affiliation{
Graduate School of Frontier Sciences, University of Tokyo, Chiba 277-5861, Japan \\
Brain Science Institute, RIKEN, Saitama 351-0198, Japan \\
PRESTO, Japan Sciencs and Technology Agency, Chiba 277-8561, Japan
}

\date{\today}

\begin{abstract}

We investigate the detection dynamics of the parallel interference canceller (PIC) for code-division multiple-access (CDMA) multiuser detection, 
applied to a randomly spread, fully syncronous base-band uncoded CDMA channel model with additive white Gaussian noise (AWGN) under perfect power control in the large-system limit. 
It is known that the predictions of the density evolution (DE) can fairly explain the detection dynamics only in the case where the detection dynamics converge. 
At transients, though, the predictions of DE systematically deviate from computer simulation results. 
Furthermore, when the detection dynamics fail to convergence, the deviation of the predictions of DE from the results of numerical experiments becomes large. 
As an alternative, generating functional analysis (GFA) can take into account the effect of the Onsager reaction term exactly and does not need the Gaussian assumption of the local field. 
We present GFA to evaluate the detection dynamics of PIC for CDMA multiuser detection. 
The predictions of GFA exhibits good consistency with the computer simulation result for any condition, even if the dynamics fail to converge. 

\end{abstract}

\maketitle

%\pacs{89.70.+c, 05.45,-a, 05.50.+q}
% PACS, the Physics and Astronomy Classification Scheme.
% http://publish.aps.org/PACS/pacsgen.html
% 89.70.+c : Information theory and communication theory
% 05.45.-a : Nonlinear dynamics and nonlinear dynamical systems (see also section 45 Classical mechanics of discrete systems)
% 05.50.+q : Lattice theory and statistics (Ising, Potts, etc.) 
%% 87.10.+e : General theory and mathematical aspects
%% 05.90.+m : Other topics in statistical physics, thermodynamics, and nonlinear dynamical systems
%% 89.90.+n : Other topics in areas of applied and interdisciplinary physics
%% 02.50.-r : Probability theory, stochastic processes, and statistics
%% 75.10.Hk : Classical spin models

\section{Introduction}
%~~~~~~~~~~~~~~~~~~~~~~~~~~~~~~~~~~~~~~~~~~~~~~~~~~~~~~~~~~~~~~~~~~~~~

Mobile communication systems, such as cellular phone systems, are now used every day by millions of people worldwide. 
Code-division multiple-access (CDMA) is a digital modulation system that employs spreading codes to enable access to a mobile communication system by multiple users \cite{Varanashi1990}. 
In the multipoint-to-point communication framework, CDMA allows several users to share a single communication channel to a base station. %<- ADDED
Each user first modulates one's own information sequence using the spreading code assigned to the user, and then the modulated sequence is transmitted to the base station. %<- ADDED
The base station receives a mixture of the transmitted signals and additional channel noise. %<- ADDED
Using the users' spreading codes, a demodulator at the base station extracts the original information sequense from the received noise-degraded mixture signal. %<- ADDED
This process is called a detection. %<- ADDED
\par %<- ADDED
Tanaka has evaluated the detection problem by the replica method \cite{Tanaka2001,Tanaka2002,Nishimori2002}. %<- ADDED
However, the detection process cannot be treated by the replica method. %<- ADDED
The detection process of CDMA has drawn much attention from theoretical as well as practical viewpoints \cite{Kabashima2003, Tanaka2005}. 
Tanaka and Okada have applied a dynamical theory of Hopfield model \cite{Okada1995} to the detection process \cite{Tanaka2005}. %<- ADDED
Their method is equivalent to the density evolution (DE) framework in the field of information theory \cite{Richadrson2001}. %<- ADDED
%Recently, the density evolution (DE) framework \cite{Okada1995, Richadrson2001} was applied to analyze some detection algorithms of the parallel interference canceller (PIC) for CDMA multiuser detection \cite{Tanaka2005}. %<- DELETED
In the DE framework, a local field, which is a matched filter output that the estimated parallel interference is subtracted from, is separated into a signal part for the detection and a remaining noise part. %<- ADDED
Furthermore, it is assumed that the noise part follows a Gaussian distribution with mean zero. %<- ADDED
%However, in order to apply density evolution to CDMA multiuser detection, it must be assumed that the local field, which is a matched filter output that the estimated parallel interference is subtracted from, follows a Gaussian distribution. %<- DELETED
%Furthermore, the Onsager reaction term included in the local field needs to be ignored in the DE framework. %<- DELETED
The predictions of DE can fairly explain the detection dynamics only in the case where the detection dynamics converge \cite{Tanaka2005}. 
However, at transients the predictions of DE systematically deviate from computer simulation results. 
%The absense of the Onsager reaction term has a more serious influence, when the detection dynamics fail to converge. %<-DELETED
The Gaussian assumption of the local field has a more serious influence, when the detection dynamics fail to converge. %<-ADDED
In such a case, the deviation of the predictions of DE from the results of numerical experiments becomes large \cite{Tanaka2005}. 
%On the other hand, generating functional analysis (GFA) \cite{Coolen2000, During1998, Kawamura2002, Mimura2004} can take into account the effect of the Onsager reaction term exactly and does not need the Gaussian assumption of the local field. %<- DELETED
On the other hand, generating functional analysis (GFA) \cite{Coolen2000, During1998, Kawamura2002, Mimura2004} does not need the Gaussian assumption. %<-ADDED
In this paper, we present GFA to evaluate the detection dynamics for CDMA multiuser detection, 
applied to a randomly spread, fully synchronous base-band uncoded CDMA channel model with additive white Gaussian noise (AWGN) under perfect power control. 
In order to confirm the validity of our analysis, we have performed computer simulations for some typical system load and channel noise conditions.

\section{System model}
%~~~~~~~~~~~~~~~~~~~~~~~~~~~~~~~~~~~~~~~~~~~~~~~~~~~~~~~~~~~~~~~~~~~~~

\begin{figure}%[htbp]
\begin{center}
\includegraphics[width=.75\linewidth,keepaspectratio]{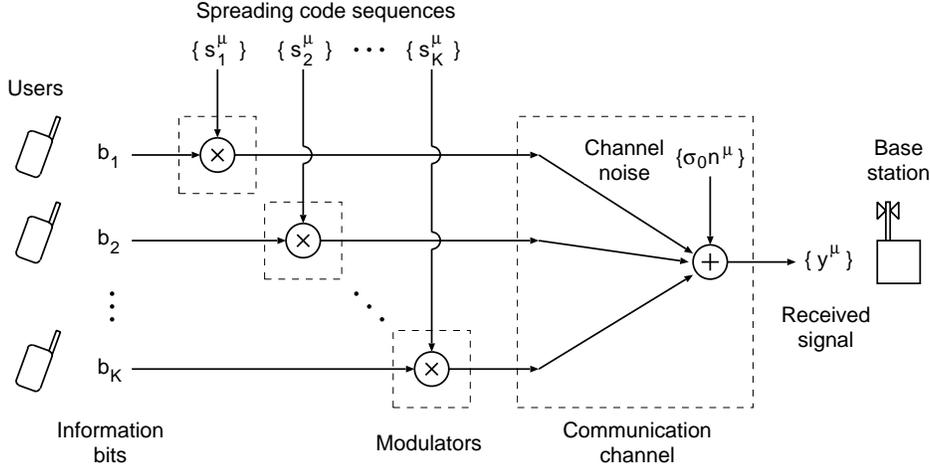}
\caption{CDMA communication model.}
\label{fig:model}
\end{center}
\end{figure}

We will focus on the basic fully syncronous $K$-user baseband binary phase-shift-keying (BPSK) CDMA channel model with perfect power control as 
\begin{equation}
y^\mu \equiv \frac1{\sqrt{N}}\sum_{k=1}^K s_k^\mu b_k + \sigma_0 n^\mu,
\end{equation}
where $y^\mu$ is the recieved signal at chip interval $\mu\in\{ 1,\cdots,N\}$, 
and where $b_k\in\{-1,1\}$ and $s_k^\mu\in\{-1,1\}$ are the BPSK-modulated information bit and the spreading code of user $k\in\{1,\cdots,K\}$ at chip interval $\mu$, respectively. 
Figure \ref{fig:model} shows this CDMA communication model. %<- ADDED
The Gaussian random variable $\sigma_0^2 n^\mu$, where $n^\mu \sim N(0,1)$, represents channel noise whose variance is $\sigma_0^2$. 
The spreading codes are independently generated from the identical unbiased distribution $P(s_k^\mu=1)=P(s_k^\mu=-1)=1/2$. 
The factor $1/\sqrt{N}$ is introduced in order to normalize the power per symbol to 1. 
Using these normalisations, the signal to noise ratio is defined as $Eb/N_0=1/(2\sigma_0^2)$. 
The ratio $\beta\equiv K/N$ is called system load. 

The goal of multiuser detection is to simultaneously infer the information bits $b_1,\cdots,b_K$ after recieving the base-band signals $y_1,\cdots,y_N$. 
The updating rule for the tentative decision $\hat{b}_k(t)\in\{-1,1\}$ of bit signal $b_k$ at stage $t$ is 
\begin{equation}
\hat{b}_k(t) = \sgn \biggl( h_k - \sum_{k' =1, \ne k}^K W_{kk'}\hat{b}_{k'}(t-1) \biggr),
\label{eq:deterministic_updating_rule}
\end{equation}
where $h_k$ is the output of the matched filter for user $k$: 
\begin{equation}
h_k \equiv \frac1{\sqrt{N}}\sum_{\mu=1}^N s_k^\mu y^\mu,
\end{equation}
and $W_{kk'}$ is the $kk'$-element of the sample correlation matrix $\vec{W}$ of the spreading code: 
\begin{equation}
W_{kk'} \equiv \frac1N \sum_{\mu=1}^N s_k^\mu s_{k'}^\mu.
\end{equation}
The function $\sgn(x)$ denotes the sign function taking 1 for $x \ge 0$ and -1 for $x<0$. 
This iterative detection algorithm is called the parallel interference canceller (PIC) \cite{Varanashi1990}. 
As for initialization, we assume the matched filter stage, i.e., $\hat{b}_k(0) = \sgn (h_k)$. 
This initialization is easily treated by formally assuming 
\begin{equation}
\hat{b}_k(-1) = 0,
\label{eq:initial_state}
\end{equation}
for all $k$. 
The widely used measure of the performance of a demodulator is the bit error rate (BER) $P_b(t)$, which is given by $P_b(t)=[1-m(t)]/2$, 
where $m(t)=\frac1K\sum_{k=1}^Kb_k\hat{b}_k(t)$ is the overlap between the information bits vector $\vec{b}(t)={}^\dagger(b_1,\cdots,b_K)$ and the tentative decision vector $\hat{\vec{b}}={}^\dagger(\hat{b}_1(t),\cdots,\hat{b}_K(t))$. 
The operator ${}^\dagger$ denotes the transpose. 
Without loss of generality, we assume that the true information bits are all 1, i.e., $b_k=1$ for all $k$, because the spreading codes are unbiased.

\section{Generating functional analysis}
%~~~~~~~~~~~~~~~~~~~~~~~~~~~~~~~~~~~~~~~~~~~~~~~~~~~~~~~~~~~~~~~~~~~~~
\subsection{Generating functional}

We analyse the detection dynamics in the large system limit where $K,N\to\infty$, while the system load $\beta$ is kept finite. 
For generating functional analysis, we introduce inverse temperature $\gamma$. 
The stochastic updating rule for the tentative decision $\hat{b}_k(t)\in\{-1,1\}$ of bit signal $b_k$ at stage $t$ is given by 
\begin{equation}
P[\hat{b}_k(t+1)=-\hat{b}_k(t)]=\frac 12 \biggl( 1-\tanh \gamma \hat{b}_k(t+1) u_k(t) \biggr),
\label{eq:updating_rule}
\end{equation}
where 
\begin{equation}
u_k(t) \equiv h_k - \sum_{k'=1, \ne k}^K W_{kk'} \hat{b}_{k'}(t) + \theta_k(t),
\end{equation}
which is called a local field. 
In the limit where $\gamma\to\infty$, this updating rule is equivalent to (\ref{eq:deterministic_updating_rule}). 
The term $\theta_k(t)$ is a time-dependent external field which is introduced in order to define a response function. 
The inverse temperature and the external field are set $\gamma\to\infty$ and $\theta_k(t)=0$ in the end of analysis. 

To analyse the dection dynamics of the system we define a generating functional $Z[\vec{\psi}]$: 
\begin{equation}
Z[\vec{\psi}]=\sum_{\hat{\vec{b}}(-1),\cdots,\hat{\vec{b}}(t)} p[\hat{\vec{b}}(-1),\cdots,\hat{\vec{b}}(t)] e^{-i\sum_{s=-1}^t\hat{\vec{b}}(s)\cdot\vec{\psi}(s)}
\label{eq:def_Z}
\end{equation}
where $\hat{\vec{b}}(s)={}^\dagger(\hat{b}_1(s),\cdots,\hat{b}_K(s))$, $\vec{\psi}(s)={}^\dagger(\psi_1(s),\cdots,\psi_K(s))$. 
In familiar way \cite{Coolen2000, During1998, Kawamura2002, Mimura2004}, one can obtain from $Z[\vec{\psi}]$ all averages of interest by differenriation, e.g., 
\begin{eqnarray}
m_k(s)&=&<\hat{b}_k(s)>=i\lim_{\vec{\psi}\to\vec{0}}\frac{\partial Z[\vec{\psi}]}{\partial \psi_k(s)}, \\
C_{kk'}(s,s')&=&<\hat{b}_k(s)\hat{b}_{k'}(s')>=-\lim_{\vec{\psi}\to\vec{0}}\frac{\partial Z[\vec{\psi}]}{\partial \psi_k(s) \partial \psi_{k'}(s')}, \\
G_{kk'}(s,s')&=&\frac{\partial <\hat{b}_k(s)>}{\partial \theta_{k'}(s')}=i\lim_{\vec{\psi}\to\vec{0}}\frac{\partial Z[\vec{\psi}]}{\partial \psi_k(s) \partial \theta_{k'}(s')}.
\end{eqnarray}
The dynamics (\ref{eq:updating_rule}) is a Markov chain, so the path probability $p[\hat{\vec{b}}(-1),\cdots,\hat{\vec{b}}(t)]$ are simply given by products of the individual transiton probabilities $\rho[\hat{\vec{b}}(s+1)|\hat{\vec{b}}(s)]$ of the chain: 
\begin{equation}
p[\hat{\vec{b}}(-1),\cdots,\hat{\vec{b}}(t)]=p[\hat{\vec{b}}(-1)]\prod_{s=-1}^{t-1}\rho[\hat{\vec{b}}(s+1)|\hat{\vec{b}}(s)], 
\label{eq:def_path_probability}
\end{equation}
where these transition probabilities are given by 
\begin{equation}
\rho[\hat{\vec{b}}(s+1)|\hat{\vec{b}}(s)]=\prod_{k=1}^K \frac{e^{\gamma\hat{b}_k(s+1)u_k(s)}}{2\cosh \gamma u_k(s)}.
\label{eq:def_transition_probability}
\end{equation}
Since the initial state is given by (\ref{eq:initial_state}), the initial state probability becomes $p[\hat{\vec{b}}(-1)=\vec{0}]=\prod_{k=1}^K p[\hat{b}_k(-1)=0]=1$. 
We separate the local field at any stage by inserting a following delta-distributions: 
\begin{equation}
1 = \int \delta\vec{u}\delta\hat{\vec{u}} \prod_{s=-1}^{t-1} \prod_{k=1}^K e^{i\hat{u}_k(s) [ u_k(s) - h_k + \sum_{k'\ne k}^K W_{kk'} \hat{b}_{k'}(s) - \theta_k(s) ] },
\label{eq:local_field}
\end{equation}
where $\delta\vec{u} \equiv \prod_{s=-1}^{t-1} \prod_{k=1}^K \frac{du_k(s)}{\sqrt{2\pi}}$ and $\delta\hat{\vec{u}} \equiv \prod_{s=-1}^{t-1} \prod_{k=1}^K \frac{d\hat{u}_k(s)}{\sqrt{2\pi}}$. 
We can express (\ref{eq:def_Z}) as 
\begin{eqnarray}
Z[\vec{\psi}]
&=& \sum_{\hat{\vec{b}}(-1),\cdots,\hat{\vec{b}}(t)} p[\hat{\vec{b}}(-1)] \int \delta\vec{u}\delta\hat{\vec{u}} \nonumber \\
& & \quad \times e^{i \sum_{s=-1}^{t-1} \sum_{k=1}^K \hat{u}_k(s) \{ u_k(s)-\hat{b}_k(s)-\theta_k(s) \} -i\sum_{s=-1}^t \sum_{k=1}^K \hat{b}_k(s)\psi_k(s)} \nonumber \\
& & \quad \times e^{\sum_{s=0}^t \sum_{k=1}^K \{ \gamma \hat{b}_k(s) u_k(s) - \ln 2 \cosh \gamma u_k(s-1) \} } \nonumber \\
& & \quad \times e^{- i \sqrt{\beta} \sigma_0 \sum_{\mu=1}^N \sum_{s=-1}^{t-1} [ \frac1{\sqrt{K}} \sum_{k=1}^K \hat{u}_k(s)s_k^\mu ] n^\mu } \nonumber \\
& & \quad \times e^{- i \beta \sum_{\mu=1}^N \sum_{s=-1}^{t-1} [ \frac1{\sqrt{K}} \sum_{k=1}^K \hat{u}_k(s)s_k^\mu ] [ \frac1{\sqrt{K}} \sum_{k'=1}^K s_{k'}^\mu \{ 1-\hat{b}_{k'}(s) \} ] }. 
\label{eq:Z1}
\end{eqnarray}
In order to average the generating functional with respect to the disorder $\{s_k^\mu\}$ and $\{n^\mu\}$,
we isolate the spreading codes by introducing the variables $v_\mu(s), w_\mu(s)$: 
\begin{eqnarray}
1 &=& \int \delta\vec{v}\delta\hat{\vec{v}} \prod_{s=-1}^{t-1} \prod_{\mu=1}^N e^{i\hat{v}_\mu(s) [ v_\mu(s) - \frac1{\sqrt{K}} \sum_{k=1}^K s_k^\mu \hat{u}_k(s) ] }, \\
1 &=& \int \delta\vec{w}\delta\hat{\vec{w}} \prod_{s=-1}^{t-1} \prod_{\mu=1}^N e^{i\hat{w}_\mu(s) [ w_\mu(s) - \frac1{\sqrt{K}} \sum_{k=1}^K s_k^\mu \{ 1-\hat{b}_k(s) \} ] },
\end{eqnarray}
where
$\delta\vec{v} \equiv \prod_{\mu=1}^{N} \prod_{s=-1}^{t-1} \frac{dv_\mu(s)}{\sqrt{2\pi}}$, 
$\delta\hat{\vec{v}} \equiv \prod_{\mu=1}^{N} \prod_{s=-1}^{t-1} \frac{d\hat{v}_\mu(s)}{\sqrt{2\pi}}$, 
$\delta\vec{w} \equiv \prod_{\mu=1}^{N} \prod_{s=-1}^{t-1} \frac{dw_\mu(s)}{\sqrt{2\pi}}$, and 
$\delta\hat{\vec{w}} \equiv \prod_{\mu=1}^{N} \prod_{s=-1}^{t-1} \frac{d\hat{w}_\mu(s)}{\sqrt{2\pi}}$. 
The term in (\ref{eq:Z1}) containing the disorder becomes 
\begin{eqnarray}
& & \overline{e^{- i \sqrt{\beta} \sigma_0 \sum_{\mu=1}^N \sum_{s=-1}^{t-1} [ \frac1{\sqrt{K}} \sum_{k=1}^K \hat{u}_k(s)s_k^\mu ] n^\mu }} \nonumber \\
& & \quad \overline{ \times e^{- i \beta \sum_{\mu=1}^N \sum_{s=-1}^{t-1} [ \frac1{\sqrt{K}} \sum_{k=1}^K \hat{u}_k(s)s_k^\mu ] [ \frac1{\sqrt{K}} \sum_{k'=1}^K s_{k'}^\mu \{ 1-\hat{b}_{k'}(s) \} ] } } \nonumber \\
&=& \int \delta\vec{v}\delta\hat{\vec{v}}\delta\vec{w}\delta\hat{\vec{w}} e^{i \sum_{\mu=1}^N \sum_{s=-1}^{t-1} \{\hat{v}_\mu(s)v_\mu(s)+\hat{w}_\mu(s)w_\mu(s)-\beta v_\mu(s)w_\mu(s)\} } \nonumber \\
& & \quad \times e^{-\frac12 \sum_{\mu=1}^N \sum_{s=-1}^{t-1} \sum_{s'=-1}^{t-1} \beta\sigma_0^2v_\mu(s)v_\mu(s') } \nonumber \\
& & \quad \times e^{-\frac12 \sum_{\mu=1}^N \sum_{s=-1}^{t-1} \sum_{s'=-1}^{t-1} \hat{v}_\mu(s) [ \frac 1K \sum_{k=1}^K \hat{u}_k(s) \hat{u}_k(s') ] \hat{v}_\mu(s') } \nonumber \\
& & \quad \times e^{-\frac12 \sum_{\mu=1}^N \sum_{s=-1}^{t-1} \sum_{s'=-1}^{t-1} \hat{v}_\mu(s) [ \frac 1K \sum_{k=1}^K \hat{u}_k(s) - \frac 1K \sum_{k=1}^K \hat{u}_k(s)\hat{b}_k(s') ] \hat{w}_\mu(s') } \nonumber \\
& & \quad \times e^{-\frac12 \sum_{\mu=1}^N \sum_{s=-1}^{t-1} \sum_{s'=-1}^{t-1} \hat{w}_\mu(s) [ \frac 1K \sum_{k=1}^K \hat{u}_k(s') - \frac 1K \sum_{k=1}^K \hat{u}_k(s')\hat{b}_k(s) ] \hat{v}_\mu(s') } \nonumber \\
& & \quad \times e^{-\frac12 \sum_{\mu=1}^N \sum_{s=-1}^{t-1} \sum_{s'=-1}^{t-1} \hat{w}_\mu(s) [ 1 -\frac 1K \sum_{k=1}^K \hat{b}_k(s) -\frac 1K \sum_{k=1}^K \hat{b}_k(s') - \frac 1K \sum_{k=1}^K \hat{b}_k(s)\hat{b}_k(s') ] \hat{w}_\mu(s') }, 
\label{eq:Z2}
\end{eqnarray}
where $\overline{\cdots}$ denotes averaging over the disorder $\{s_k^\mu\}$ and $\{n^\mu\}$. 
We separate the relevant one-stage and two-stage order parameters by inserting: 
\begin{eqnarray}
1 &=& \biggl(\frac K{2\pi}\biggr)^{t+1} \int d\vec{m}d\hat{\vec{m}} e^{i K \sum_{s=-1}^{t-1} \hat{m}(s) [ m(s) - \frac1{\sqrt{K}} \sum_{k=1}^K \hat{b}_k(s) ] }, \\
1 &=& \biggl(\frac K{2\pi}\biggr)^{t+1} \int d\vec{k}d\hat{\vec{k}} e^{i K \sum_{s=-1}^{t-1} \hat{k}(s) [ k(s) - \frac1{\sqrt{K}} \sum_{k=1}^K \hat{u}_k(s) ] }, \\
1 &=& \biggl(\frac K{2\pi}\biggr)^{(t+1)^2} \int d\vec{q}d\hat{\vec{q}} e^{i K \sum_{s=-1}^{t-1} \sum_{s'=-1}^{t-1} \hat{q}(s,s') [ q(s,s') - \frac1{\sqrt{K}} \sum_{k=1}^K \hat{b}_k(s) \hat{b}_k(s') ] }, \\
1 &=& \biggl(\frac K{2\pi}\biggr)^{(t+1)^2} \int d\vec{Q}d\hat{\vec{Q}} e^{i K \sum_{s=-1}^{t-1} \sum_{s'=-1}^{t-1} \hat{Q}(s,s') [ Q(s,s') - \frac1{\sqrt{K}} \sum_{k=1}^K \hat{u}_k(s) \hat{u}_k(s') ] }, \\
1 &=& \biggl(\frac K{2\pi}\biggr)^{(t+1)^2} \int d\vec{L}d\hat{\vec{L}} e^{i K \sum_{s=-1}^{t-1} \sum_{s'=-1}^{t-1} \hat{L}(s,s') [ L(s,s') - \frac1{\sqrt{K}} \sum_{k=1}^K \hat{b}_k(s) \hat{u}_k(s') ] }.
\end{eqnarray}
Since the initial state probability is factorisable, the disorder-averaged generating functional factorises into single-site contributions. 
The disorder-averaged generating functional is for $K\to\infty$ dominated by a saddle-point. 
We can thus simplify the saddle-point problem to 
\begin{equation}
\bar{Z}[\vec{\psi}] = \int 
d\vec{m}d\hat{\vec{m}}
d\vec{k}d\hat{\vec{k}}
d\vec{q}d\hat{\vec{q}}
d\vec{Q}d\hat{\vec{Q}}
d\vec{L}d\hat{\vec{L}}
e^{K(\Psi+\Phi+\Omega)+O(\ln K)}
\label{eq:Z3}
\end{equation}
in which the functions $\Psi, \Phi, \Omega$ are given by 
\begin{eqnarray}
\Psi
&\equiv& i \sum_{s=-1}^{t-1} \{ \hat{m}(s) m(s) + \hat{k}(s)k(s) \} \nonumber \\
& & + i \sum_{s=-1}^{t-1} \sum_{s'=0}^{t-1} \{ \hat{q}(s,s') q(s,s') + \hat{Q}(s,s') Q(s,s') + \hat{L}(s,s') L(s,s') \} \\
\Phi
&\equiv& \frac 1K \sum_{k=1}^K \ln \biggl\{ \sum_{\hat{b}(-1),\cdots,\hat{b}(t)} p[\hat{b}(-1)] \int \delta u \delta\hat{u} e^{\sum_{s=0}^t \{ \gamma \hat{b}(s) u(s-1) - \ln 2 \cosh \gamma u(s-1) \} } \nonumber \\
& & \quad \times e^{- i \sum_{s=-1}^{t-1} \sum_{s'=-1}^{t-1} \{ \hat{q}(s,s') \hat{b}(s) \hat{b}(s') + \hat{Q}(s,s') \hat{u}(s) \hat{u}(s') + \hat{L}(s,s') \hat{b}(s) \hat{u}(s') \} } \nonumber \\
& & \quad \times e^{i \sum_{s=-1}^{t-1} \hat{u}(s) \{ u(s) - \hat{b}(s) - \theta_k(s) - \hat{k}(s) \} - i \sum_{s=-1}^{t-1} \hat{b}(s) \hat{m}(s) - i \sum_{s=-1}^t \hat{b}(s) \psi_k(s) } \biggr\} \\
\Omega
&\equiv& \frac 1K \ln \int \delta\vec{v}\delta\hat{\vec{v}}\delta\vec{w}\delta\hat{\vec{w}} e^{i \sum_{\mu=1}^N \sum_{s=-1}^{t-1} \{ \hat{v}_\mu(s) v_\mu(s) + \hat{w}_\mu(s) w_\mu(s) - \beta v_\mu(s) w_\mu(s) \} } \nonumber \\
& & \quad \times e^{- \frac 12 \sum_{\mu=1}^N \sum_{s=-1}^{t-1} \sum_{s'=-1}^{t-1} \{ \beta \sigma_0^2 v_\mu(s) v_\mu(s') + \hat{v}_\mu(s) Q(s,s') \hat{v}_\mu(s') \} } \nonumber \\
& & \quad \times e^{- \frac 12 \sum_{\mu=1}^N \sum_{s=-1}^{t-1} \sum_{s'=-1}^{t-1} \{ \hat{v}_\mu(s) [k(s) - L(s',s)] \hat{w}_\mu(s') + \hat{w}_\mu(s) [k(s') - L(s,s')] \hat{v}_\mu(s') \} }\nonumber \\
& & \quad \times e^{- \frac 12 \sum_{\mu=1}^N \sum_{s=-1}^{t-1} \sum_{s'=-1}^{t-1} \{ \hat{w}_\mu(s) [1 - m(s) - m(s') + q(s,s')] \hat{w}_\mu(s') \} }
\end{eqnarray}
where $\delta u \equiv \prod_{s=0}^{t-1} \frac{du(s)}{\sqrt{2\pi}}$ and $\delta \hat{u} \equiv \prod_{s=0}^{t-1} \frac{d\hat{u}(s)}{\sqrt{2\pi}}$. 
We have arrived at a single-site saddle-point problem. 
Using normalization condition and $\bar{Z}[\vec{0}]=1$, we obtain field derivatives of the generating functional as follows: 
\begin{eqnarray}
%& &\ll\hat{u}(s)\gg=\ll\hat{u}(s)\hat{u}(s')\gg =0, \\
& & \overline{<\hat{b}_k(s)>}=<\hat{b}(s)>_*, \\
& & \overline{<\hat{b}_k(s)\hat{b}_{k'}(s')>}=\delta_{k,k'}<\hat{b}(s)\hat{b}(s')>_* +(1-\delta_{k,k'})<\hat{b}(s)>_*<\hat{b}(s')>_*, \\
& & \frac{\partial}{\partial \theta_{k'}(s')}\overline{<\hat{b}_k(s)>}=-i\delta_{k,k'}<\hat{b}(s)\hat{u}(s')>_*, 
\end{eqnarray}
where $\delta_{k,k'}$ is Kronecker's delta taking 1 if $k=k'$ and 0 otherwise and $<\cdot>_*$ denotes 
\begin{equation}
<f(\{\hat{b},u,\hat{u}\})>_* \equiv \frac{\sum_{\hat{b}(-1),\cdots,\hat{b}(t)} \int \delta u\delta \hat{u} M(\{\hat{b},u,\hat{u}\}) f(\{\hat{b},u,\hat{u}\})}{\sum_{\hat{b}(-1),\cdots,\hat{b}(t)} \int \delta u\delta \hat{u} M(\{\hat{b},u,\hat{u}\})}, \label{eq:<>*}
\end{equation}
with 
\begin{eqnarray}
M(\{\hat{b},u,\hat{u}\}) 
&\equiv& p[\hat{b}(-1)] e^{\sum_{s=0}^t \{ \gamma \hat{b}(s) u(s-1) - \ln 2 \cosh \gamma u(s-1) \} } \nonumber \\
& & \quad \times e^{- i \sum_{s=-1}^{t-1} \sum_{s'=-1}^{t-1} \{ \hat{q}(s,s') \hat{b}(s) \hat{b}(s') + \hat{Q}(s,s') \hat{u}(s) \hat{u}(s') + \hat{L}(s,s') \hat{b}(s) \hat{u}(s') \} } \nonumber \\
& & \quad \times e^{i \sum_{s=-1}^{t-1} \hat{u}(s) \{ u(s) - \hat{b}(s) - \theta_k(s) - \hat{k}(s) \} - i \sum_{s=-1}^{t-1} \hat{b}(s) \hat{m}(s) } |_{saddle}. 
\end{eqnarray}
The evaluation $f|_{saddle}$ denotes an evaluation of a function $f$ at the dominating saddle point. 
Therefore we see the order parameters are essentially single-site ones.

\subsection{saddle-point equations}
%~~~~~~~~~~~~~~~~~~~~~~~~~~~~~~~~~~~~~~~~~~~~~~~~~~~~~~~~~~~~~~~~~~~~~

In the limit $K\to\infty$, the integral (\ref{eq:Z3}) will be dominated by the saddle point of the extensive exponent $\Psi+\Phi+\Omega$. 
We first calculate the remaining Gaussian integral in $\Omega$: 
\begin{eqnarray}
\Omega
&=& \frac 1\beta \int \frac{d\hat{\vec{v}}}{(2\pi)^{(t+1)/2}} \frac{d\hat{\vec{w}}}{(2\pi)^{(t+1)/2}} e^{
i {}^\dagger \hat{\vec{w}} (\beta^{-1} \vec{1}) \hat{\vec{v}} 
- \frac12 {}^\dagger \hat{\vec{v}} \vec{Q} \hat{\vec{v}}
- \frac12 {}^\dagger \hat{\vec{v}} {}^\dagger \vec{B} \hat{\vec{w}}
- \frac12 {}^\dagger \hat{\vec{w}} \vec{B} \hat{\vec{v}}
- \frac12 {}^\dagger \hat{\vec{w}} \hat{\vec{D}} \hat{\vec{w}} } \nonumber \\
&=& \frac 1\beta \int \frac{d\hat{\vec{v}}}{(2\pi)^{t/2}} e^{- \frac12 {}^\dagger \hat{\vec{v}} \vec{Q} \hat{\vec{v}}} |\hat{\vec{D}}|^{-1/2}e^{ -\frac12 {}^\dagger \hat{\vec{v}} {}^\dagger (\beta^{-1} \vec{1} -\vec{B}) \hat{\vec{D}}^{-1} (\beta^{-1} \vec{1} -\vec{B}) \hat{\vec{v}}} \nonumber \\
&=& -\frac 1{2\beta} \biggl( \ln |\hat{\vec{D}}| + \ln |\vec{Q}+ {}^\dagger (\beta^{-1} \vec{1} -\vec{B}) \hat{\vec{D}}^{-1} (\beta^{-1} \vec{1} -\vec{B})| \biggr),
\end{eqnarray}
where $\vec{B}$, $\hat{\vec{D}}$ and $\vec{Q}$ are matrices having matrix elements 
\begin{eqnarray}
B(s,s') &\equiv& -ik(s')-G(s,s'), \\
\hat{D}(s,s') &\equiv& \frac{\sigma_0^2}\beta + 1-m(s)-m(s')+C(s,s'),
\end{eqnarray}
and $Q(s,s')$, respectively.
The saddle-point equations are derived by differentiation with respect to integration variables $\{\vec{m},\hat{\vec{m}},\vec{k},\hat{\vec{k}},\vec{q},\hat{\vec{q}},\vec{Q},\hat{\vec{Q}},\vec{L},\hat{\vec{L}}\}$. 
These equations will involve the average single-site correlation $C(s,s')$ and the average single-site response function $G(s,s')$: 
\begin{eqnarray}
%m(s)&=&\lim_{K\to\infty}\frac1K\sum_{k=1}^K \overline{<\hat{b}_k(s)>}=\ll \hat{b}(s) \gg, \\
C(s,s')&=&\lim_{K\to\infty}\frac1K\sum_{k=1}^K \overline{<\hat{b}_k(s)\hat{b}_{k'}(s')>}=<\hat{b}(s)\hat{b}(s')>_*, \\
G(s,s')&=&\lim_{K\to\infty}\frac1K\sum_{k=1}^K \frac{\partial}{\partial \theta_{k'}(s')}\overline{<\hat{b}_k(s)>}=-i<\hat{b}(s)\hat{u}(s')>_* .
\end{eqnarray}
Straightforward differentiation by usage of causality, leads us to the following saddle-point equations: 
\begin{eqnarray}
\hat{m}(s) &=& k(s) = \hat{q}(s,s') = Q(s,s') = 0, \label{eq:sp1} \\
\hat{k}(s) &=& |\vec{\Lambda}_s|, \label{eq:final_khat} \\
m(s) &=& <\hat{b}(s)>_*, \\
q(s,s') &=& <\hat{b}(s)\hat{b}(s')>_*= C(s,s'), \\
L(s,s') &=& i G(s,s') = \left\{
\begin{array}{ll}
-i <\hat{b}(s)\hat{u}(s')>_*, & {\rm for} \; s>s' \\
0, & {\rm for} \; s\le s',
\end{array}
\right. \\
\hat{\vec{Q}} &=& -i\frac 12 {}^\dagger (\vec{1}+\beta \vec{G})^{-1} \vec{D} (\vec{1}+\beta {}^\dagger \vec{G})^{-1}, \\
\hat{\vec{L}} &=& (\vec{1}-\beta{}^\dagger \vec{G})^{-1}, \label{eq:sp2}
\end{eqnarray}
where $\hat{\vec{Q}}$, $\hat{\vec{L}}$, $\vec{D}$ and $\vec{\Lambda}_s$ are matrices having matrix elements $\hat{Q}(s,s')$, $\hat{L}(s,s')$, 
\begin{equation}
D(s,s') \equiv \beta \hat{D}(s,s') = \sigma_0^2 + \beta [ 1-m(s)-m(s')+C(s,s') ], 
\end{equation}
and 
\begin{equation}
\Lambda_s (s',s'') \equiv
\left\{
\begin{array}{ll}
\delta_{s',s''} + \beta G(s'',s'), & {\rm for} \; s'\ne s \\
1, & {\rm for} \; s' = s, \\
\end{array}
\right. 
\end{equation}
respectively. 
Substituting (\ref{eq:sp1})-(\ref{eq:sp2}) into (\ref{eq:<>*}) and introducing a simple rescaling of local fields and conjugate local fields, the term $<\hat{b}(s)\hat{\vec{u}}>_*$ becomes 
\begin{eqnarray}
<\hat{b}(s)\hat{\vec{u}}>_*
&=& \int \delta u \delta \hat{u} \sum_{\hat{b}(-1),\cdots,\hat{b}(t)} \hat{b}(s) \hat{\vec{u}} e^{\sum_{s=-1}^{t-1} \{ \gamma \hat{b}(s+1) u(s) - \ln 2 \cosh \gamma u(s) \} } \nonumber \\
& & \quad \times e^{-\frac12{}^\dagger\hat{\vec{u}}\vec{R}\hat{\vec{u}}+i\hat{\vec{u}}(\vec{u}-\hat{\vec{k}}-\vec{\theta}-\vec{\Gamma} \hat{\vec{b}})} \nonumber \\
&=& i \int \frac{d\vec{v}e^{-\frac 12\vec{v}\cdot R^{-1}\vec{v}}}{\sqrt{|2\pi \vec{R}|}} \sum_{\hat{b}(-1),\cdots,\hat{b}(t)} \hat{b}(s)\vec{R}^{-1}\vec{v} \prod_{s=-1}^{t-1} \frac 12 [1+\hat{b}(s+1)\sgn u(s)], 
\end{eqnarray}
in the limit $\gamma\to\infty$, where $\hat{\vec{u}}\equiv{}^\dagger (\hat{u}(-1),\cdots,\hat{u}(t-1))$, $\vec{R}\equiv(\vec{1}+\beta {}^\dagger\vec{G})^{-1} \vec{D} (\vec{1}+\beta \vec{G})^{-1}$ and $\vec{\Gamma}\equiv(\vec{1}+\beta \vec{G})^{-1}\beta \vec{G}$. 
The terms $<\hat{b}(s)>_*$ and $<\hat{b}(s)\hat{b}(s')>_*$ can also be calculated in a similar way. 

Let us summarize our calculation. 
Some macroscopic integration variables are found to vanish in the relevant physical saddle-point: $\hat{m(s)} = k(s) = \hat{q}(s,s') = Q(s,s') = 0$. 
The remainig ones can all be expressed in terms of three macroscopic observables, namely the overlaps $m(s)$, the single-site correlation functions $C(s,s')$ and the single-site response functions $G(s,s')$. 
Finally, setting $\gamma\to\infty$ and $\theta(s)=0$, we then arrive at the following saddle-point equations in the thermodynamic limit, i.e., $K\to\infty$: 
\begin{eqnarray}
m(s)&=&\ll \hat{b}(s) \gg, \label{eq:sp_m} \\
C(s,s')&=&\ll \hat{b}(s) \hat{b}(s') \gg, \\
G(s,s')&=&
\left\{
\begin{array}{ll}
\ll \hat{b}(s)(\vec{R}^{-1}\vec{v})(s') \gg, & {\rm for} \; s>s' \\
0, & {\rm for} \; s \le s'.
\end{array}
\right. 
\end{eqnarray}
The bit error rate is obtained by 
\begin{equation}
P_b(s)=\frac{1-m(s)}2.
\end{equation}
The average over the effective path measure is given by 
\begin{eqnarray}
\ll g(\hat{\vec{b},\vec{v}}) \gg &\equiv& \int {\cal D}\vec{v} \; {\rm Tr} \; g(\hat{\vec{b}},\vec{v}) \prod_{s=-1}^{t-1} \frac 12 [1+\hat{b}(s+1)\sgn u(s)], \\
{\cal D}\vec{v} &\equiv& \frac{d\vec{v}e^{-\frac 12\vec{v}\cdot \vec{R}^{-1}\vec{v}}}{\sqrt{|2\pi \vec{R}|}}, \\
{\rm Tr} &\equiv& \sum_{\hat{b}(-1)\in\{0\}, \hat{b}(0),\cdots,\hat{b}(t)\in\{ -1,1\} }, \\
u(s)&=&\hat{k}(s)+v(s)+(\vec{\Gamma} \hat{\vec{b}})(s), \label{eq:final_local_field} \\
\vec{R} &=& (\vec{1}+\beta {}^\dagger \vec{G})^{-1} \vec{D} (\vec{1}+\beta \vec{G})^{-1}, \\
\vec{\Gamma} &=& (\vec{1}+\beta \vec{G})^{-1}\beta \vec{G}, \\
\hat{k}(s) &=& |\vec{\Lambda}_s|, \\
D(s,s') &\equiv& \sigma_0^2 + \beta [ 1-m(s)-m(s')+C(s,s') ], \\
\Lambda_s(s',s'') &=&
\left\{
\begin{array}{ll}
\delta_{s',s''} + \beta G(s'',s'), & {\rm for} \; s'\ne s \\
1, & {\rm for} \; s' = s. \\
\end{array}
\right. \label{eq:sp_Lambda}
\end{eqnarray}
The terms $(\vec{R}^{-1}\vec{v})(s)$ and $(\vec{\Gamma} \vec{\sigma})(s)$ denote the $s$th element of the vector $\vec{R}^{-1}\vec{v}$ and $\vec{\Gamma} \vec{\sigma}$,respectively. 
Equations (\ref{eq:sp_m})-(\ref{eq:sp_Lambda}) entirely describe the dynamics of the system. 
In the limit where $t\to\infty$, the term $(\vec{\Gamma} \vec{\sigma})(s)$ in (\ref{eq:final_local_field}) can be regarded as a self-interaction and corresponds to the Onsager reaction term in equilibrium statistical mechanics. 
Therefore, in this paper, we call this term the Onsager reaction term.

\section{Results and discussion}
%~~~~~~~~~~~~~~~~~~~~~~~~~~~~~~~~~~~~~~~~~~~~~~~~~~~~~~~~~~~~~~~~~~~~~

\begin{figure}%[htbp]
\begin{center}
\includegraphics[width=.6\linewidth,keepaspectratio]{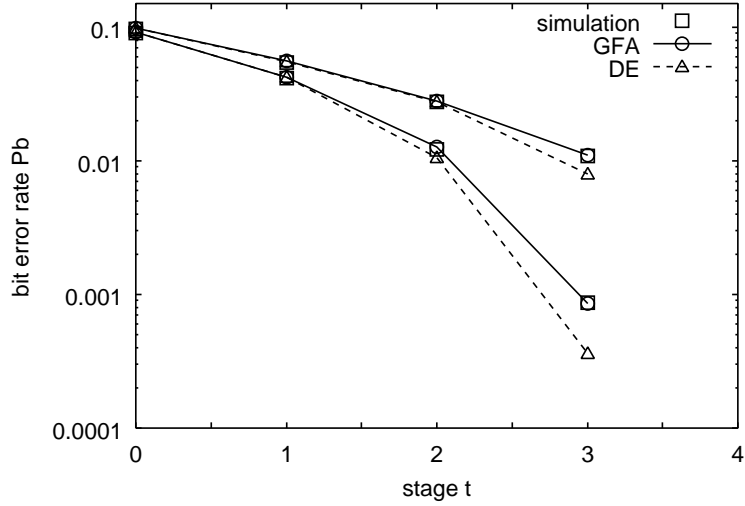}
\caption{
The first few stages of the detection dynamics predicted by generating functional analysis (solid line) and density evolution (dashed line). 
Computer simulations (square) are evaluated with $N=8000$ from 100 experiments for the cases $E_b/N_0=$ 7.0 [dB] (upper), 9.0 [dB] (lower). 
The system load is $\beta=0.5<\beta_c$ for both cases. 
}
\label{fig:Beta=05}
\end{center}
\end{figure}

\begin{figure}%[htbp]
\begin{center}
\includegraphics[width=.6\linewidth,keepaspectratio]{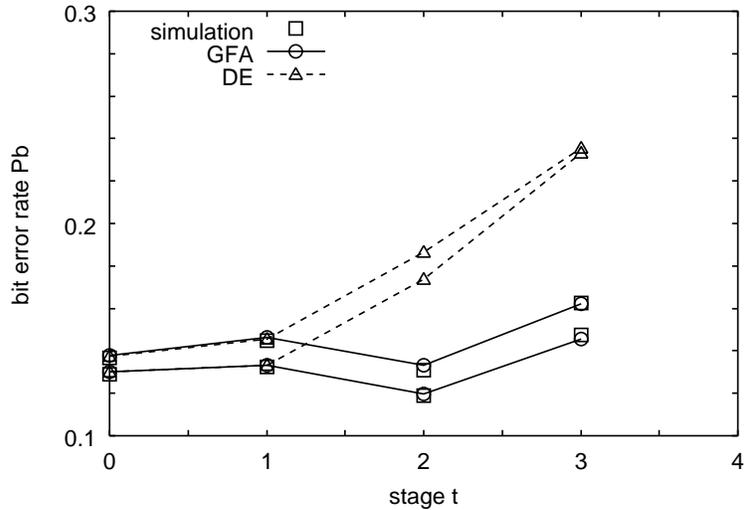}
\caption{
The first few stages of the detection dynamics predicted by generating functional analysis (solid line) and density evolution (dashed line). 
Computer simulations (square) are evaluated with $N=8000$ from 100 experiments for the cases $E_b/N_0=$ 5.5 [dB] (upper), 7.5 [dB] (lower). 
The system load is $\beta=0.7>\beta_c$ for both cases. 
}
\label{fig:Beta=07}
\end{center}
\end{figure}

In order to validate the results obtained above, 
we performed numerical experiments in an $N=8000$ system. 
Figure \ref{fig:Beta=05} shows the first few stages of the detection dynamics obtained from 100 experiments for the cases $E_b/N_0=$ 7.0, 9.0 [dB], 
predicted by generating functional analysis (GFA) and density evolution (DE) \cite{Tanaka2005}, where $E_b/N_0$ [dB] denotes $10 \log_{10} E_b/N_0$ (see Appendix \ref{app:DE}). 
The system load is $\beta=0.5<\beta_c$, where $\beta_c$ is the critical system load defined as the minimum system load at which the dynamics fail to convergence. 
Figure \ref{fig:Beta=07} shows the first few stages of the detection dynamics obtained from 100 experiments for the cases $E_b/N_0=$ 5.5, 7.5 [dB], predicted by GFA and DE with the system load $\beta=0.7>\beta_c$. 
Oscillation of the detection dynamics was observed, when $\beta>\beta_c$. 
In such a case, both GFA and DE predicted the failure of convergence of the dynamics. 
However, the DE results has residual deviations in figures \ref{fig:Beta=05} and \ref{fig:Beta=07} due to the lack of the Onsager reaction term and the assumption that the local field follows a Gaussian distribution. 
In particular, the deviation of the DE predictions from the simulation results becomes large when $\beta>\beta_c$. 
In contrast, GFA exhibits good consistency with the simulation results for any system load. 

The difference between DE and GFA appears also in a signal term with respect to the information bit of the local field. 
The signal terms of DE and GFA at stage $t$ represent $B_t$ and $\hat{k}(t)$, respectively (see Appendix \ref{app:DE}). 
The signal term $\hat{k}(t)$ derived by GFA contains all response functions $G(s,s')$ with $s,s'\le t$. 
On the other hand, the signal term $B_t$ derived by DE contains only the response functions of adjacent stages. 
This difference appears from stage $t=1$. 
The signal term $\hat{k}(1)$ of GFA is, 
\begin{eqnarray}
\hat{k}(1)
&=&
\left|
\begin{array}{ccc}
1 & \beta G(0,-1) & \beta G(1,-1) \\
0 & 1 & \beta G(1,0) \\
1 & 1 & 1
\end{array}
\right| \nonumber \\
&=& 1 - \beta G(1,0) + \beta^2 G(1,0) G(0,-1)-\beta G(1,-1),
\end{eqnarray}
while the signal term $B_1$ of DE is 
\begin{equation}
B_1 = 1 - \beta U_1 +\beta^2 U_1 U_0. 
\end{equation}
As you can easily see, the $B_1$ contains only $U_1$ and $U_0$, which correspond to $G(1,0)$ and $G(0,-1)$ of GFA respectively, 
while the $\hat{k}(1)$ has the response function between stage 1 and stage -1 as $G(1,-1)$.


\section{Conclusions}
%~~~~~~~~~~~~~~~~~~~~~~~~~~~~~~~~~~~~~~~~~~~~~~~~~~~~~~~~~~~~~~~~~~~~~

We presented the generating functional analysis to describe the detection dynamics of PIC for CDMA multiuser detection. 
The predictions of DE can qualitatively explain the detection dynamics only when the detection dynamics converge. 
Furthermore, the deviation of the predictions of DE from the results of numerical experiments becomes large when the detection dynamics fail to convergence. 
In contrast, the predictions of GFA are in good agreement with computer simulation result of PIC for any system load and channel noise level, even if the dynamics fail to converge.

\begin{flushleft}
{\small \bf Acknowledgement}
\end{flushleft}
%~~~~~~~~~~~~~~~~~~~~~~~~~~~~~~~~~~~~~~~~~~~~~~~~~~~~~~~~~~~~~~~~~~~~~

This work was partially supported by a Grant-in-Aid 
for Scientific Research on Priority Areas No. 14084212, 
and for Scientific Research (C) No. 16500093 
from the Ministry of Education, Culture, Sports, Science and Technology of Japan.

\appendix

\section{Density evolution of CDMA detection dynamics\label{app:DE}}
%~~~~~~~~~~~~~~~~~~~~~~~~~~~~~~~~~~~~~~~~~~~~~~~~~~~~~~~~~~~~~~~~~~~~~

Density evolution is a useful tool to analyze nonlinear dynamics \cite{Okada1995,Richadrson2001}. 
By means of density evolution, the bit error rate $P_b(t)$ of hard decisions $\hat{b}_k(t)=\sgn[u_k(t-1)]$ at the $t$th stage is given by 
\begin{equation}
P_b(t) = \frac{1-M_t}2, 
\end{equation}
where $M_t$ are to be evaluated by the following recursive formulas for $B_t$, $C_{t,\tau}$, $M_t$, $U_t$ and $q_{t,\tau}$: 
\begin{eqnarray}
B_t &=& 1-\beta U_t B_{t-1}, \\
C_{t,\tau} &=& V_{t,\tau}+\beta^2 U_t U_\tau C_{t-1,\tau-1} \nonumber \\
           & & \quad +\sum_{\lambda=-1}^{t-1}V_{\lambda,\tau} \prod_{\kappa=\lambda+1}^t (-\beta U_\kappa)+\sum_{\lambda=-1}^{\tau-1}V_{\lambda,t} \prod_{\kappa=\lambda+1}^\tau (-\beta U_\kappa), \\
V_{t,\tau}&=& \sigma_0^2 + \beta (1-M_t-M_\tau+q_{t,\tau}), \\
M_{t+1}&=&\int Dz \sgn(B_t+z\sqrt{C_{t,t}}), \\
U_{t+1}&=&\frac1{\sqrt{C_{t,t}}}\int Dz \;z\sgn(B_t+z\sqrt{C_{t,t}}), \\
q_{t+1,\tau+1}&=&\int \int \int Dz Du Dv \sgn (B_t+z\sqrt{C_{t,\tau}}+u\sqrt{C_{t,t}-C_{t,\tau}}) \nonumber \\
              & & \quad \times \sgn (B_\tau+z\sqrt{C_{t,\tau}}+v\sqrt{C_{t,t}-C_{t,\tau}}), 
\end{eqnarray}
where $Dz\equiv (2\pi)^{-1/2}e^{-z^2/2}dz$. 
The initializations are 
$V_{-1,t}=V_{t,-1}=\sigma_0^2+\beta(1-M_t)$, 
$B_{-1}=1$, 
$M_{-1}=0$, 
$C_{-1,-1}=\sigma_0^2+\beta$, 
$C_{-1,t}=C_{t,-1}=V_{-1,t}-\beta U_t V_{-1,t-1}$, 
and 
$q_{-1,t}=q_{t,-1}=0$. 
The physical meaning of the parameters $B_t$, $C_{t,\tau}$, $M_t$, $U_t$ and $q_{t,\tau}$ is 
\begin{eqnarray}
B_t&=&{\rm E} [u_k(t)], \\
C_{t,\tau}&=&{\rm Cov} [u_k(t),u_k(\tau)], \\
M_{t+1}&=&\frac1K\sum_{k=1}^K\sgn[u_k(t)], \\
U_{t+1}&=&\frac1K\sum_{k=1}^K\sgn'[u_k(t)], \\
q_{t+1,\tau+1}&=&\frac1K\sum_{k=1}^K\sgn[u_k(t)]\sgn[u_k(\tau)]. 
\end{eqnarray}
The detailed derivation is available in the appendix of the paper \cite{Tanaka2005}. 
In the derivation by means of density evolution, it is assumed that the local field $u_k(t)$ follows the Gaussian distribution with mean $B_t$ and covariance $C_{t,\tau}$. 
Furthermore, the Onsager reaction term is ignored. 
The signal term $B_t$ contains only the response functions of adjacent stages.

\end{document}